\documentclass[conference]{IEEEtran}

\pdfoutput=1

\usepackage[numbers,compress]{natbib}

\usepackage{microtype}   

\usepackage[keeplastbox]{flushend} 

\usepackage{graphicx}
\usepackage{epstopdf}
\DeclareGraphicsExtensions{.eps,.pdf}
\usepackage{calc}

\usepackage[table]{xcolor}
\usepackage{multirow}

\let\labelindent\relax
\usepackage{enumitem}

\usepackage{url}

\usepackage{pgf}
\usepackage{pgfplots}
\usepackage{tikz}
\usetikzlibrary{arrows,automata,positioning}

\usepackage{fancyhdr}
\fancypagestyle{plain}{
\fancyhf{}
\fancyhead[L]{2020 CSI/CPSSI International Symposium on Real-Time and Embedded Systems and Technologies (RTEST)}
\fancyfoot[L]{978-1-7281-7551-5/20/\$31.00 \copyright{}2020 IEEE}

}
\newcommand*\circled[1]{\tikz[baseline=(char.base)]{
            \node[shape=circle,draw,inner sep=1pt,scale=0.9] (char) {#1};}}

\begin{document}
\bstctlcite{IEEEexample:BSTcontrol}

\title{Software-Based Monitoring and Analysis of a \\USB Host Controller Subject to \\Electrostatic Discharge}

\author{%
	\IEEEauthorblockN{%
		Natasha Jarus,
		Antonio Sabatini,
		Pratik Maheshwari,
		and Sahra Sedigh Sarvestani
	}
	\IEEEauthorblockA{%
		Department of Electrical and Computer Engineering\\
		Missouri University of Science and Technology, Rolla, MO 65409, USA\\
		Email: \{jarus, ajs5gd, prm8c7, sedighs\}@mst.edu
	}
}

\maketitle
\thispagestyle{plain} 

\begin{abstract}
	Observing, understanding, and mitigating the effects of failure in embedded systems is essential for building dependable control systems.
	We develop a software-based monitoring methodology to further this goal.
	This methodology can be applied to any embedded system peripheral and allows the system to operate normally while the monitoring software is running.
	We use software to instrument the operating system kernel and record indicators of system behavior.
	By comparing those indicators against baseline indicators of normal system operation, faults can be detected and appropriate action can be taken.

	We implement this methodology to detect faults caused by electrostatic discharge in a USB host controller.
	As indicators, we select specific control registers that provide a manifestation of the internal execution of the host controller.
	Analysis of the recorded register values reveals differences in system execution when the system is subject to interference.
	This improved understanding of system behavior may lead to better hardware and software mitigation of electrostatic discharge and assist in root-cause analysis and repair of failures.
\end{abstract}


\IEEEpeerreviewmaketitle

\section{Introduction}

As embedded systems become smaller and smaller, they become more vulnerable to physical events and thus more difficult to make reliable.
Interference from Electrostatic Discharge (ESD) is a major cause of this unreliability, since a smaller electrical charge is required for smaller components to experience an ESD event.
The effects of these events on the software running on the embedded system are not yet well understood.
In order to understand these effects, we must observe how the hardware effects of ESD manifest in the software controlling that hardware.

Depending on severity, ESD events can cause permanent hardware damage or manifest as software glitches, such as screen flickers or program crashes, that appear random and unexpected to the system user.
Associating these user-observed failures with specific software and hardware faults is an ongoing challenge.
Additionally, component miniaturization increases the difficulty of monitoring all traces on a board for ESD with hardware probes.
Even if all the traces could be monitored, it is a nontrivial task to analyze where the ESD coupled to the system and the resulting effect it had on various components.
Finally, while invasive hardware testing might be feasible on a development board, testing on consumer hardware not equipped with test points and monitoring hardware is much more difficult.
Executing in-field tests or analyzing faults that only occur on production hardware are daunting tasks.

We propose a low-level, lightweight software-based method for monitoring, detecting, and analyzing the effects of ESD events.
Our method is applicable to other types of electromagnetic interference, but in the interest of clarity, the focus of this paper is on ESD events.
Software instrumentation allows for monitoring of hardware that cannot be physically probed.
Some existing software analysis techniques focus on high-level failures, e.g., screen glitches, but stop short of root-cause analysis of hardware faults.
Other software approaches study low-level failures, such as data corruption in CPU caches, but require complete control of the system unmediated by an operating system and are thus inapplicable to systems under typical usage conditions.
Our approach uses modified hardware drivers to allow a system in the field to be monitored for ESD events.

With software instrumentation, we are able to observe changes in system operation caused by ESD\@.
We compare this to operation during normal system operation to determine whether a system is experiencing an ESD event.
These results have several applications in failure analysis as well as hardware and software design.
Collected data can be used for postmortem analysis, validating system designs, and runtime fault detection and recovery.
Throughout this paper, we will discuss these observations and analyses within the context of a USB host controller on an embedded system, specifically, an ARM system running Linux.
In our work, we consider small-scale ESD events that do not persist after a power cycle.

In this paper, we present:
\begin{itemize}
	\item A method for instrumenting device drivers to monitor internal operation of system peripherals.
	\item A method for analyzing the observed states of peripheral operation.
\end{itemize}
\IEEEpubidadjcol

The rest of the paper is as follows: Section~\ref{sec:related-work} reviews related literature.
Our software instrumentation approach is described in Section~\ref{sec:monitoring}.
Section~\ref{sec:analysis} discusses the data analysis algorithm.
Experimental setup is documented in Section~\ref{sec:hardware} and results and observations are presented in Section~\ref{sec:results}.
Finally, Section~\ref{sec:conclusion} draws conclusions and discusses future extensions to this work.

\section{Background and Related Literature}
\label{sec:related-work}


ESD-induced failures can be broadly categorized as either \emph{hard failures} or \emph{soft failures}~\cite{industry_council_on_esd_target_levels_white_2019}.
In this context, a hard failure permanently damages the system so that components must be replaced.
Soft failures, on the other hand, can be recovered from; these failures are further characterized into three levels based on the visibility of the failure and the action needed to recover from it:
\begin{enumerate}[start=1, label={Level \arabic*)}, leftmargin=*, labelindent=\parindent]
	\item The system automatically recovers with no user-visible faults or loss or corruption of data.
		Often this recovery is possible due to ESD-robust hardware and fault-tolerant control protocols.
	\item The system experiences a system-level manifestation, such as momentary screen or data corruption, but recovers without intervention.
	\item The system crashes or requires the user to perform an action, such as resetting the system or unplugging and re-plugging a device, to recover from a fault condition.
\end{enumerate}
These failures are studied using a variety of hardware- and software-based techniques.

Numerous studies have investigated the relationship between ESD interference and level 2 and 3 soft failures.
Hardware ESD fault injection with direct injection and field injection probes is described in \cite{MaL11,KiY11,ScO13}.  
These studies characterize integrated circuit (IC) immunity to ESD\@.
The sensitivity threshold for each IC was determined by injecting ESD at increasing voltages and observing when errors occurred.
In these studies, only user-visible errors, such as screen glitches or hardware resets, were investigated.

\citet{izadi_systematic_2018} extend this fault injection process by mapping the ESD sensitivity of the board.
The injection probes are attached to a 2-D scanner that sweeps them across the board.
At each point on a grid over the CPU, ESD is injected and the level at which the device becomes susceptible is recorded.
The resulting map can be used to identify traces and components that are at risk for ESD damage.
Mapping is carried out at various CPU loads and clock speeds; the authors determine that the system is most susceptible under heavy load and low clock speed.


\citet{vora_application_2016} study user-visible soft failures in a microprocessor, a microcontroller, and an {FPGA}.
In particular, they observed a relationship between CPU load and likelihood of display flicker on a microprocessor,
indicating that ESD was coupling to the CPU chip rather than to the display itself.
Furthermore, they observed that the likelihood of certain failures---process termination and display flicker---depend on the program executing at the time of the ESD event.

Investigating level 1 soft failures and understanding the root causes of higher-level soft failures requires the ability to observe a system's behavior at a high level of detail.
\citet{vora_application_2016,vora_hardware_2018,feng_guilty_2019} use a custom microcontroller running code which monitors register values and system interrupts to study the effects of ESD on CPUs.
While too invasive to use on a system performing additional tasks, this approach gives a very fine-grain view of observable soft failures.
In particular, the authors observe numerous multiple bit errors in IO registers and frequent spurious interrupt triggers.

The effect of ESD on USB devices in particular has also been investigated.
\citet{maghlakelidze_pin_2018} develop an automated testing system for studying soft failures in a USB interface on a single-board computer.
The system is characterized by injecting ESD pulses of varied voltage and pulse width into specific IC pins.
Soft failures are observed based on data transmission rate and error messages in kernel logs.
Under positive voltage injections, most failures did not require user interaction; however, negative voltage injections produced numerous severe soft failures.
\citet{koch_identification_2019} further test USB-related soft failures and determine that likelihood of failure is also dependent on the state of the USB protocol,
i.e., which packets are being transmitted at the time of the injection.
Root cause analysis shows that many failures are caused by ESD coupling to the power domains in the USB controller rather than to data lines.

While some soft failures are not user-visible, they may still be observable by software monitoring of low-level system behavior.
\citet{6341069} continuously poll the status of a phase-lock loop (PLL) embedded in the microcontroller; if the PLL unlocks, it can be assumed that the system has experienced an ESD shock.
While this approach provides an excellent measure of ESD events on the microcontroller, it cannot measure peripheral ESD events because most peripherals do not contain a separate PLL that can be monitored by the microcontroller.

Another case study of low-level system monitoring is carried out in~\cite{liu_preliminary_2018} on a wireless router.
A debugging serial port on the router logs every context switch performed by the processors, giving an approximate record of the execution path taken by processes running on the router.
This data is collected into system function graphs of both reference operating function and ESD-exposed function.
Several graph metrics are applied to these graphs; differences in metric values indicate that soft-failures can be observed by this monitoring technique.

While not directly related to ESD events, software-based as well as combined hardware and software system monitoring approaches have been studied extensively.
\citet{4365762} outline research related to monitoring for runtime verification.
System state is monitored by some combination of hardware and software; this information is then used to verify that the system is operating within specification.
A software-specific study of fault monitoring is carried out by \cite{1377185}. 
The authors present a taxonomy of runtime monitoring approaches and discuss various system requirements for different monitoring techniques.

\citet{ChoudhuriG09} develop a mixed hardware and software approach for logging non-deterministic behavior in embedded systems.
They modify a compiler to emit code that logs messages to an attached storage system, reducing processing overhead on the low-powered embedded hardware being monitored.
\citet{TeinHor12} create a tool that converts an embedded system software specification into both an executable and a configuration for a hardware monitor.
The hardware monitor interfaces with the embedded CPU and its communication buses and verifies the operation of the system.

\citet{1377185} develop a software-based monitoring system for ATMs by instrumenting the drivers for each hardware component to measure state and performance.
A runtime checker uses the resulting data to determine if the system is operating correctly.
If not, recovery actions can be taken to restore system availability.



The goal of our work is to improve the resolution of ESD software detection---in effect, to make some level 1 soft failures visible to detection software---and to better understand the software and hardware root causes of all types of soft failures.
Early detection of ESD effects and detailed logs of system behavior are essential to tracing ESD as it propagates through a system.
We aim to achieve this with minimal impact to system behavior, as the visibility and behavior of ESD-induced failures can change based on the processes running on the system.
Allowing a system to operate as it does in the field provides a better basis for testing ESD effects and reproducing unusual ESD-induced failures.
Finally, such instrumentation enables real-time monitoring and recovery from faults, improving system immunity to level 2 and 3 soft failures.

\section{Proposed Monitoring Approach}
\label{sec:monitoring}

Inducing ESD events on an embedded system peripheral causes bits to flip in its data or control lines and power glitches that corrupt computations.
These flipped bits can lead to changes in register values, loss of synchronization between peripherals and the CPU, or data corruption.
All of these effects are visible to software running on the embedded system and thus should be detectable by monitoring software.
Our objective is to use software to log as many of these events as possible for analysis.
Monitoring only for corruption of transmitted data may be confounded by protocol-level checksums and retransmissions; furthermore, doing so only observes the effects of ESD on data lines and misses ESD events caused by discharges into the chip power supplies.
The methodology we propose offers a lower-level view of peripheral operation that captures a wider array of ESD events.

\def\ohcigraphic{\includegraphics[width=\columnwidth]{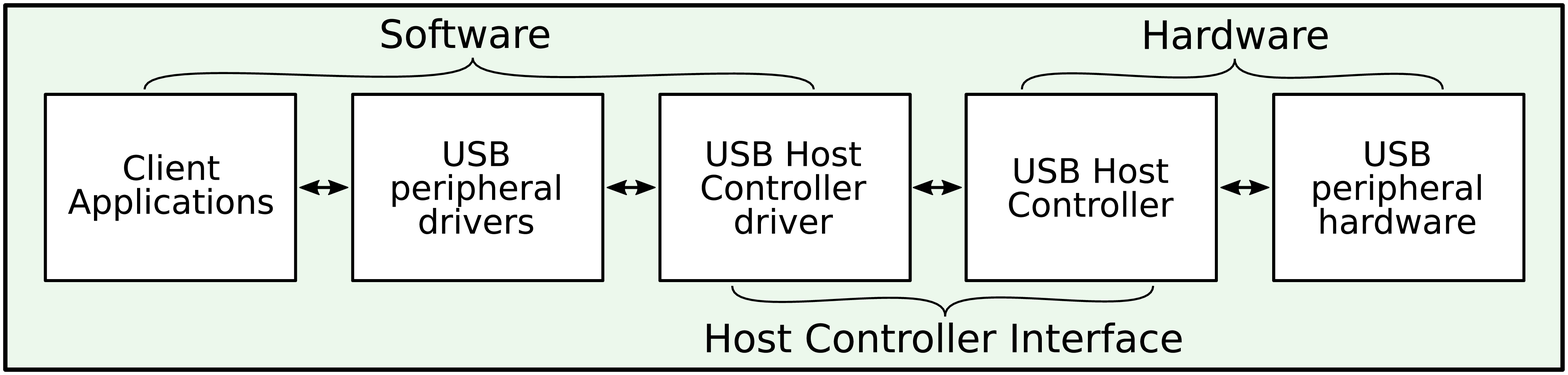}}
\newlength\ohciheight
\settoheight\ohciheight{\ohcigraphic}
\begin{figure}
	\ohcigraphic
	\caption{USB subsystem block diagram}
	\label{fig:ohci}
\end{figure}

This work can be applied to many computer peripherals, but we present it in the context of a USB Host Controller on an embedded system running Linux (see Section~\ref{sec:hardware} for more detail).
The Host Controller serves as an interface between the physical USB hardware and the software executing on the system's CPU as shown in Figure~\ref{fig:ohci}.
Its responsibilities include enumerating devices as they are connected and disconnected, configuring power delivery, and communicating data and control signals between the system's memory and the USB peripherals.
We select it for instrumentation as it connects directly to the USB bus and is thus subject to any ESD events happening on the bus.

Our work focuses on non-invasive monitoring of the effects of ESD events using software that allows normal system operation.
We primarily study changes in register values, as those values control the operation of the peripheral device. 
Each system state is represented by the $n$-tuple consisting of the values of each peripheral register at a specific time.
Some of these changes will be part of normal operation. 
When ESD is induced, however, we should observe new abnormal states or unexpected transitions between normal states.
These abnormal states and transitions can indicate that the system is experiencing ESD\@.
Our analysis avoids state-space explosion by only considering states that are observed during system operation; it does not exhaustively explore the state space.

\begin{figure}
	\centering
	\includegraphics[height=\ohciheight]{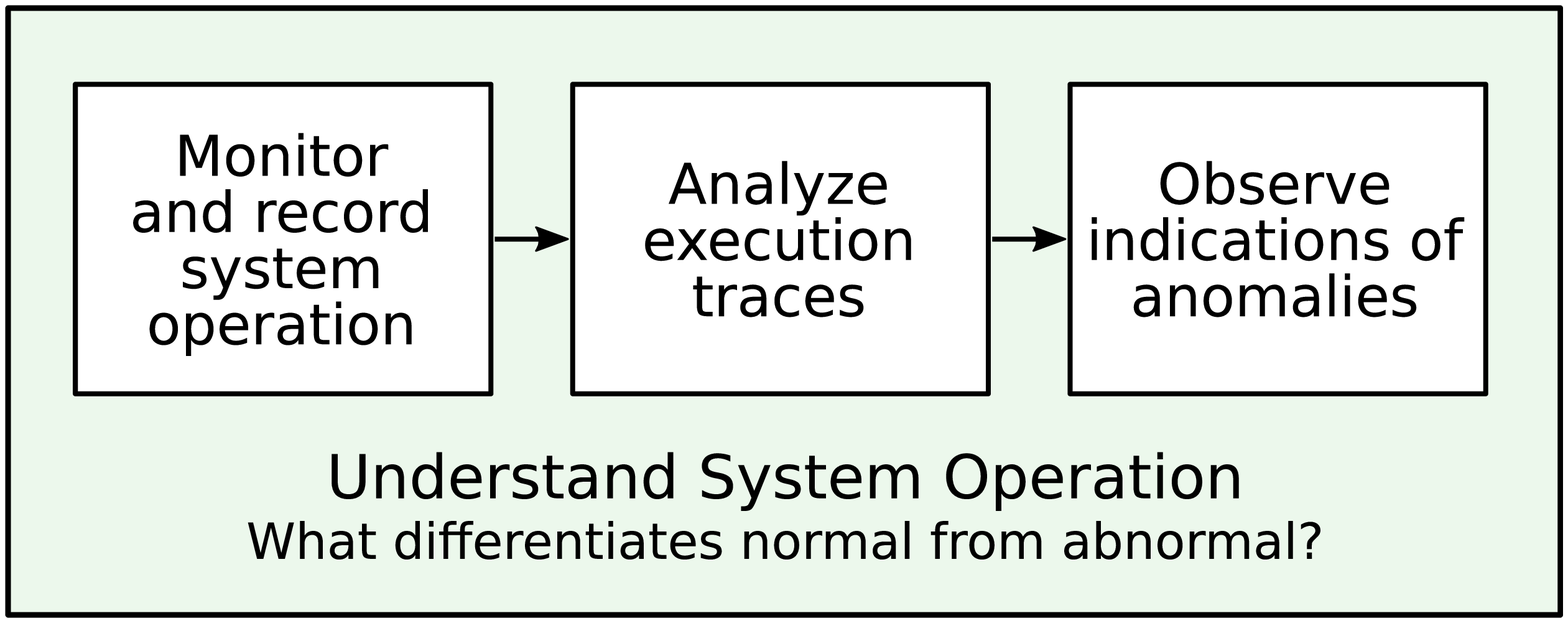}
	\caption{Research Methodology}
	\label{fig:methodology}
\end{figure}

The USB host controller is a complex piece of hardware whose operation is quite opaque to the system CPU.
We cannot inspect any of its internal registers or microcode execution process.
The extent of our visibility into its operation is the control registers it exposes to the system.
We select registers to monitor that provide a manifestation of the host controller's internal state.
Recording snapshots of register values as the system performs USB operations gives a trace of host controller execution.
The goal of this research is to use these traces to observe anomalous operation potentially caused by ESD, as summarized in Figure~\ref{fig:methodology}.

While the host controller's registers are mapped in system memory, Linux's memory protection mechanisms prevent unprivileged programs from reading them.
Thus, we must insert some software into the Linux kernel to allow us access to those memory addresses.
Initially we attempted a naive approach which repeatedly sampled the registers in a loop.
After this approach proved ineffective, we developed a more sophisticated approach which captures register values every time they are relevant to software executing on the CPU.

\subsection{Initial Design}

Our first design focused on directly reading USB register values from their physical memory addresses.
We adapted the Myregrw~\cite{myregrw} software to better suit our needs as a softprobe for ESD\@.
This software consists of a Linux system driver and a program that communicates with it.
The driver reads the values of requested physical memory addresses. 
The user-level program reads a configuration file specifying which addresses to request, repeatedly requests the data at those addresses, and stores that data to a file.

We configured Myregrw to record the control and status registers for the {USB} host interface.\footnote{These are mapped in the physical address range \texttt{0x49000000}--\texttt{0x49000014}.}
We injected ESD into the host controller while Myregrw continuously sampled the registers.
In theory, ESD-caused changes should appear in the recorded register values. 

However, the sampling rate of this softprobe was not sufficient to observe {ESD}-induced errors.
We empirically determined that the sampling rate of the software running on the system used in our experiments is, on average, 342Hz.
We can assume in the worst case that the system executes one instruction per cycle and would reset a register value after one instruction.
The system we ran these experiments on has a 400MHz clock (see Section~\ref{sec:hardware} for details).
We can calculate a pessimistic lower bound on the sampling rate by assuming the worst case scenario of a register value changing every clock cycle.
In this situation, we would log on average $\frac{342}{400*10^{6}}*100=0.000856\%$ of its values.
As we have twenty-three registers to monitor, the effective sampling rate will be even lower.
Considering this low probability and the lack of information recorded from our experiments, we devised a new measurement methodology with a higher sampling rate capable of recording additional register values.

A confounding issue with this approach is the competition for access to these values between the Myregrw driver and the USB host controller driver.
By default, Linux drivers have execution priority over any user applications, meaning that it would be nearly impossible to read all of the register values after an error but before the USB host controller driver modifies the registers.
Therefore, we developed a new methodology that, in addition to providing a faster sampling rate, ensures the register values are recorded before the USB host controller driver can modify them.

\subsection{Improved Design}
In the improved approach, we first enabled the debugging configuration already present in the USB host controller driver.
We then modified the drivers for the {USB} host controller.
The host controller driver consists of several functions that are called when certain events occur; for example, \texttt{ohci\_irq} is called when an IRQ occurs for the host controller.
We configured each function to first log its name and the values of the host controller registers to the system log.
These modifications allow us to observe not only register state changes but also the order in which different driver functions are called. 
An example of such a log entry is shown in Figure~\ref{fig:hc-state}.

This approach is minimally invasive as the driver modifications are minor and do not affect the logic of the driver itself.
While this induces a constant overhead, in practice the overhead is small\footnote{Estimated less than 10\% overhead.} and can be reduced by using, e.g., a buffer to hold log entries and a separate program to write those entries to a file.
The sampling rate is variable, but it exactly captures the driver-visible operation of the USB host controller.

\begin{figure}
	\scriptsize
	\begin{verbatim}
	function: ohci_irq
	HcControl: 0x83
	HcCommandStatus: 0x4
	HcInterruptStatus: 0x24
	HcInterruptEnable: 0x8000005e
	HcInterruptDisable: 0x8000005e
	HcHCCA: 0x338b1000
	HcPeriodCurrentED: 0x0
	HcControlHeadED: 0x339b2000
	HcControlCurrentED: 0x0
	HcBulkHeadED: 0x339b2080
	HcBulkCurrentED: 0x0
	HcDoneHead: 0x0
	HcFmInterval: 0xa7782edf
	HcFmRemaining: 0x80002760
	HcFmNumber: 0x921d
	HcPeriodicStart: 0x2a2f
	HcLSThreshold: 0x628
	HcRhDescriptorA: 0x2001202
	HcRhDescriptorB: 0x0
	HcRhStatus: 0x8000
	HcRhDescriptorA: 0x2001202
	HcRhPortStatus[0]: 0x103
	HcRhPortStatus[1]: 0x100
	Done.
	\end{verbatim}
	\caption{Example host controller state}
	\label{fig:hc-state}
\end{figure}

\section{Proposed Analysis Approach}
\label{sec:analysis}

The log files generated by the instrumented driver consist of lines each having a timestamp, register name, and associated register value. 
We parse these lines into $n$-tuples containing snapshots of register values at the time of each function call.
The sequence of $n$-tuples from each log constitutes an execution trace.

Many of these execution traces revisit the same states repeatedly.
By identifying these repeated states and coalescing them, we can develop an execution graph.
This graph is a directed graph where each node is a unique system state and an edge from node $s$ to node $t$ indicates that the system went from state $s$ to state $t$ in the corresponding execution trace.
An execution trace then becomes a path through the execution graph.

Once execution graphs for each log have been created, we repeat the deduplication process to produce the unified execution graph of all runs.
This allows us to identify similarities and differences among system execution traces.

The operation of the analysis code can be summarized as follows:
\begin{enumerate}
	\item Parse the log files to create states based on the registers' values.
	\item Deduplicate these execution traces to derive a per-run execution graph.
	\item Deduplicate the execution graphs of different runs to derive a universal execution graph.
	\item Using this execution graph and each run's execution trace, perform statistical analysis on the data.
\end{enumerate}

\subsection{Constructing Execution Graphs}

The first stage of analysis parses register values from the log file for each run.
After creating tuples for each of the states in that file, we deduplicate the sequence of states to create the nodes of the execution graph.
We then derive the execution trace path through the execution graph from the state sequence.
We also record the number of times each transition is taken.

\subsection{Constructing the Unified Execution Graph}

The next analysis step combines the data from each log into a unified execution graph.
The process is similar to that used to develop the execution graph for each log.
Certain registers for the host controller contain memory addresses that change every time the driver is reloaded.\footnote{%
These registers are \texttt{HcPeriodCurrentED}, \texttt{HcBulkCurrentED}, \texttt{HcFmRemaining}, \texttt{HcHCCA}, \texttt{HcControlHeadED}, \texttt{HcControlCurrentED}, \texttt{HcBulkHeadED}, \texttt{HcFmNumber}, and \texttt{HcDoneHead}.}
The values of these registers are not significant to our analysis; however, changes in the values are as they indicate changes in host controller execution.
Therefore, we create a new globally unique state each time the value changes within a trace.

\subsection{Graph Analysis}

We divide the data collected from test runs into two groups: baseline and ESD-exposed. 
Baseline logs are logs of the system operating normally; they provide us with the system's expected state machine.
ESD-exposed logs document how the system transitions into and out of unexpected behavior due to ESD exposure.

After we create the graph of globally unique states, we analyze the baseline and ESD-exposed logs individually to observe how system operation differs among them.
We subtract the set of states reached in baseline logs from the set of states reached in ESD-exposed logs to get a list of states only reached during {ESD} injection.
These state sets can be used to show where and when the system transitioned into a state that can potentially be attributed to {ESD}.
Similarly, we can determine which transitions between states are present only during {ESD} exposure.

\vspace*{-1.5ex}
\section{Case Study}
\label{sec:hardware}

The system used for tests was the FriendlyArm Mini2440 embedded development platform with a Samsung S3C2440 ARM926T processor \cite{FALinux}.
Its {USB} host interface conforms to the Open Host Controller Interface specifications~\cite{ohci}.
The system ran a modified Linux kernel based on the version 2.6.29 kernel downloaded from the FriendlyArm website \cite{FALinux}.\footnote{We have successfully replicated this study on a more recent Linux kernel version; the results are forthcoming.}
We set up the system with our logging software and connected it to a PC to control it during the tests.
During testing, a standard USB 2.0 flash drive was connected to the system's USB port.
To ensure that the host controller is active during ESD injection, we copied a large file to or from the flash drive during tests.

To thoroughly characterize system behavior, {ESD} interference was injected using electric (E) field and magnetic (H) field probes powered by a transmission line pulse (TLP) generator.
For each probe, multiple tests were run with varying pulse voltages.
In addition, different sizes of probes were used to adjust the intensity of the fields injected.
The E-field probe does not have an orientation; we positioned it across the {USB} port or over the host controller IC\@.
E-field interference was injected using an EZ-3 probe at voltages between 500 and 5500 volts with a pulse width between 0.1 and 0.25 seconds.
Because the magnetic fields generated by the H-field probe are directional, we conducted tests with the probe in parallel with and perpendicular to the data and control lines.
We used two probes, the HX-5 and the HX-1T2, injecting ESD betweeen 500 and 8000 volts with pulse widths between 0.1 and 0.6 seconds.
The system was more resilient to H-field interference, allowing us to perform H-field tests with more intense ESD conditions than were possible with E-field tests.
\section{Results}
\label{sec:results}

\subsection{Registers of Interest}

Certain registers on the host controller were observed to give indications of ESD\@.
In particular, we consider the values of the registers for interrupt enabling and disabling (\texttt{HcInterruptEnable} and \texttt{HcInterruptDisable}), interrupt status (\texttt{HcInterruptStatus}), control (\texttt{HcControl}), and port status (\texttt{HcRhPortStatus0}).
The host controller has multiple events and errors it can generate hardware interrupts for; the driver can enable and disable them depending on the current operation and check whether they have been triggered via the interrupt enable, disable, and status registers.
The control register allows the driver to switch between various USB transfer modes and enable certain host controller features.
The port status register reports whether a port is enabled, what device is connected to a port, device power configuration, etc.

Per the OHCI specification~\cite{ohci}, \texttt{HcInterruptEnable} and \texttt{HcInterruptDisable} should be duplicates of each other when read.
However, as shown in Table~\ref{tbl:interrupt-enable}, there are a few states in the ESD-exposed data where they are not duplicates.
This may indicate ESD-induced bit flips or the system failing to properly update both registers when one is changed. 

\begin{table}
	\centering
	\caption{Probability Distribution of Register Values: \texttt{HcInterruptEnable} and \texttt{HcInterruptDisable}}
	\begin{tabular}{c|c|r r r}
		\multirow{2}{*}{\bfseries{Value}} & \bfseries{Baseline} & \multicolumn{3}{c}{\bfseries{ESD-exposed Probability}} \\
																			& \bfseries{Probability} & \bfseries{\texttt{Enable}} & \bfseries{\texttt{Disable}} & \bfseries{Difference} \\ \hline 
		\rowcolor{gray!25}
		0x8000005e & 0.22289 & 0.20041 & 0.20041 & 0 \\
		0x8000001a & 0.01350 & 0.09677 & 0.09666 & 0.00011 \\ 
		\rowcolor{gray!25}
		0x8000005a & 0.76156 & 0.65932 & 0.65943 & -0.00011 \\
		0x8000001e & 0.00202 & 0.04359 & 0.04359 & 0 \\
		\hline
	\end{tabular}
	\label{tbl:interrupt-enable}
\end{table}

\begin{figure}
	\centering
	\includegraphics[width=\columnwidth]{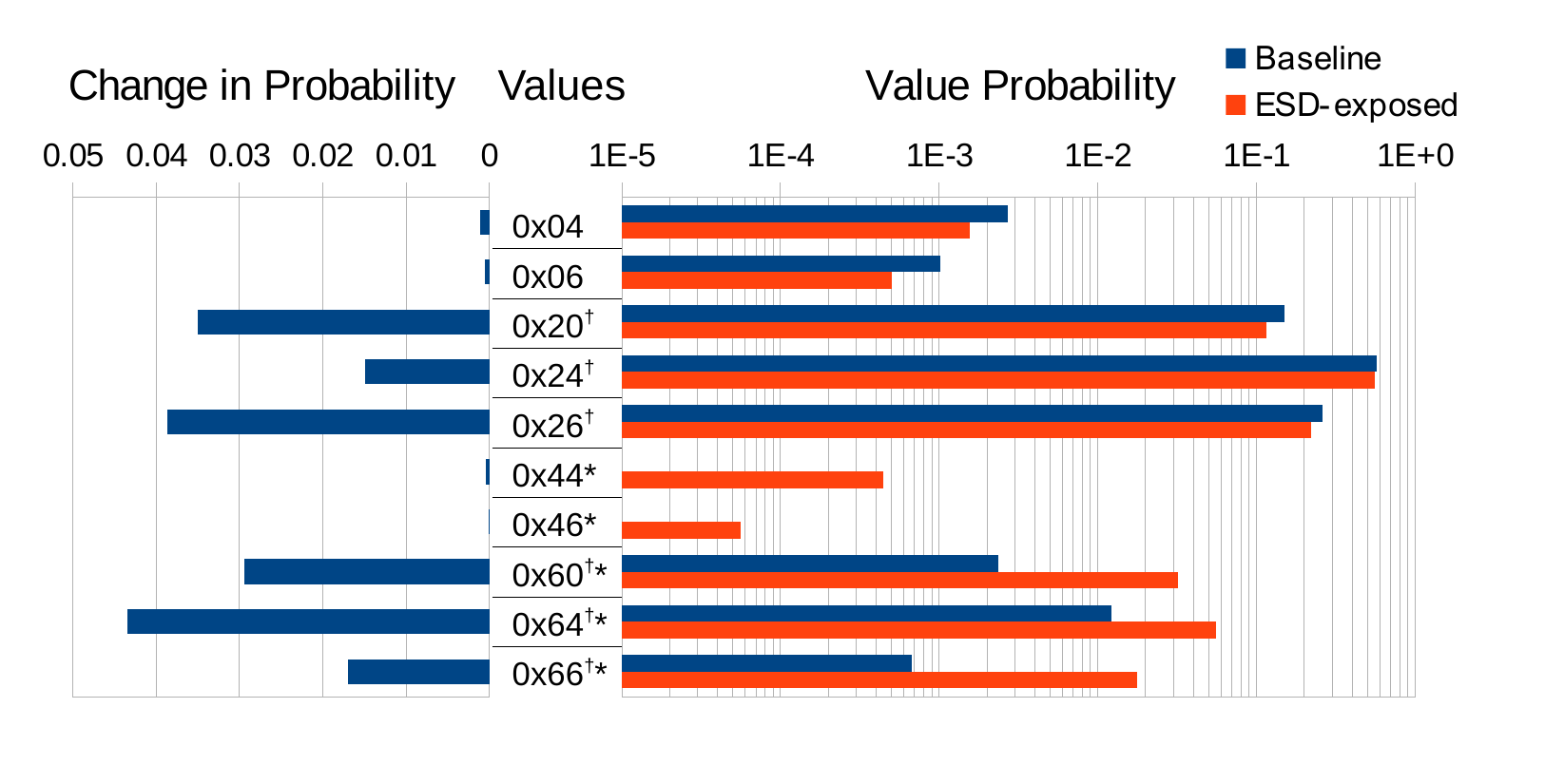}
	\caption[]{Probability Distribution of Register Values: \texttt{HcInterruptStatus} \\ $^\dag$ indicates frame counter overflow; * indicates status change}
	\label{fig:hcinterruptstatus}
\end{figure}

The \texttt{HcInterruptStatus} register values observed are shown in Figure~\ref{fig:hcinterruptstatus} along with the probability of those values appearing in baseline and ESD-exposed logs and the absolute change in that probability due to ESD exposure. 
It shows a dramatic increase in values where the frame number counter overflowed (marked $^\dag$) in the ESD-exposed logs, indicating that the system transmits many more frames during ESD exposure. 
In addition, values indicating the hub's status has changed (marked *) are also much more prevalent in ESD-exposed logs. 

\begin{figure}
	\centering
	\includegraphics[width=\columnwidth]{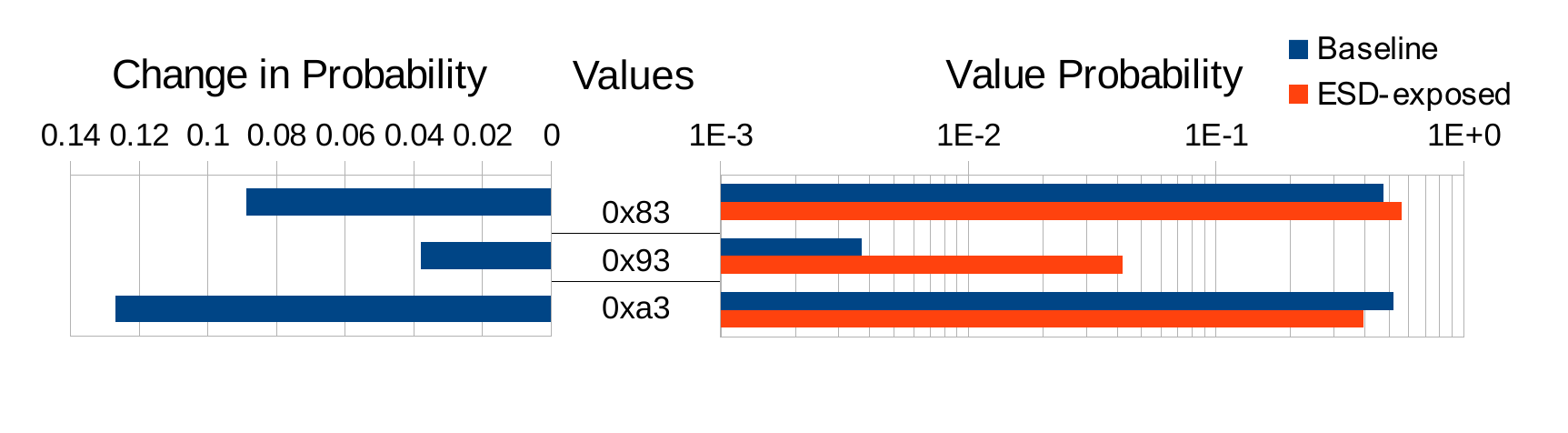}
	\vspace*{-4.7ex}
	\caption[]{Probability Distribution of Register Values: \texttt{HcControl}}
	\label{fig:hccontrol}
\end{figure}

The \texttt{HcControl} register values provide a different perspective on the increase in the number of frames and hub status changes.
Figure~\ref{fig:hccontrol} shows a great increase in control frame processing (\texttt{0x93}) and a corresponding decrease in bulk data frame processing (\texttt{0xa3}). 
It is possible that ESD glitches are disrupting bus operation, requiring the host controller and device to send a greater number of status change frames.
In addition, corruption in the bulk data frames would require retransmissions and therefore increase the number of new control and data frames (\texttt{0x83}).

\begin{figure}
	\centering
	\includegraphics[width=\columnwidth]{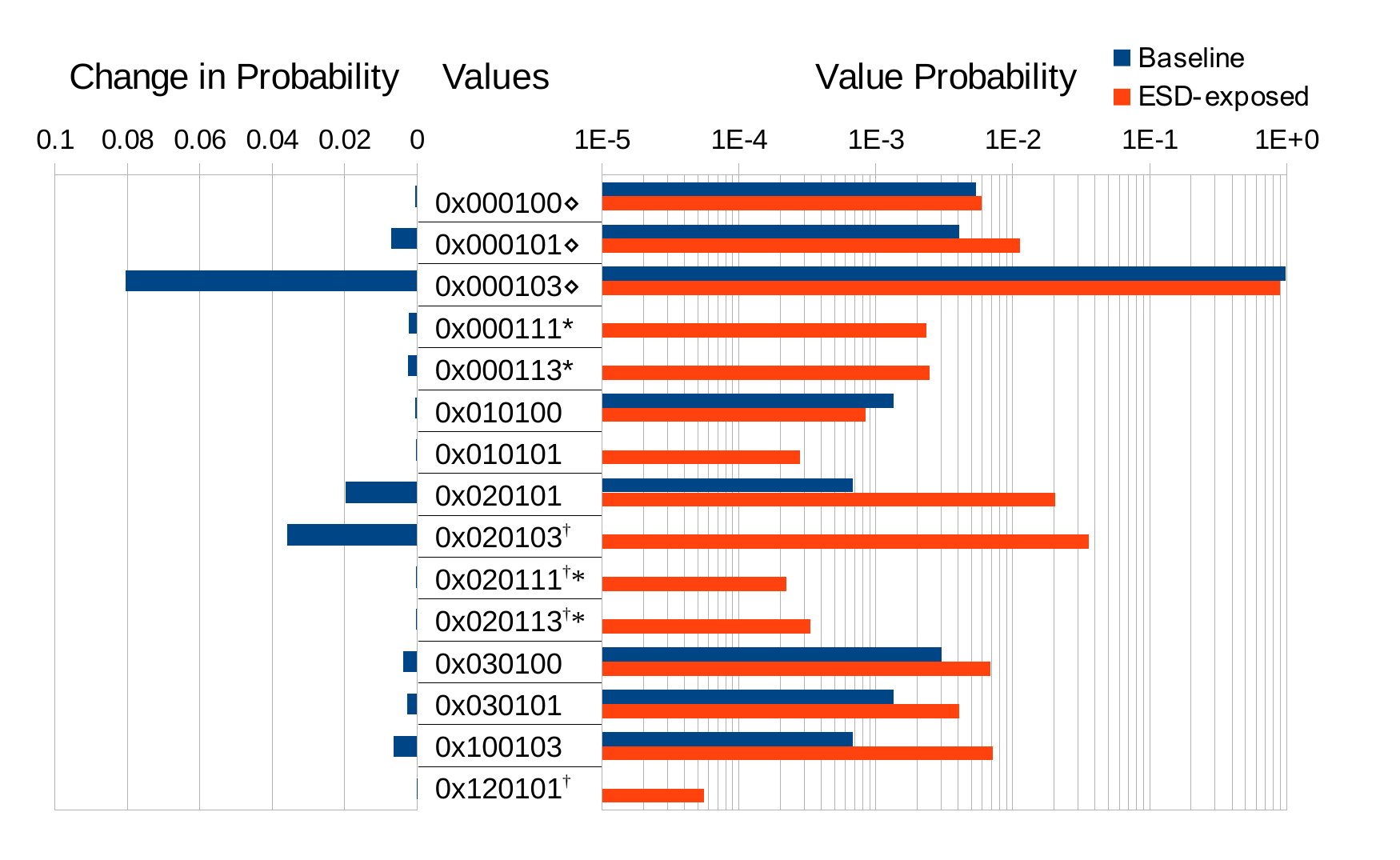}
	\caption[]{Probability Distribution of Register Values: \texttt{HcRhPortStatus0}\\ $\diamond$ indicates device connected with no change in port status; $^\dag$ indicates port enabled/disabled; * indicates port reset}
	\label{fig:hcrhportstatus0}
\end{figure}

The \texttt{HcRhPortStatus0} register contains status information about the port the USB drive was plugged into during testing.
Figure~\ref{fig:hcrhportstatus0} shows a marked decrease in states where the port status remains unchanged ($\diamond$) and an increase in states indicating the port has been enabled or disabled ($^\dag$). 
As well, port resets (*) were only observed in ESD-exposed logs. 
The prevalence of resets and toggling whether the port is enabled hint that the host controller is experiencing unexpected errors and attempting to recover by resetting the port's status. 
The presence of a port reset where the driver or host controller would not usually issue one is a particularly strong indicator of ESD exposure. 

\subsection{Execution Graphs}
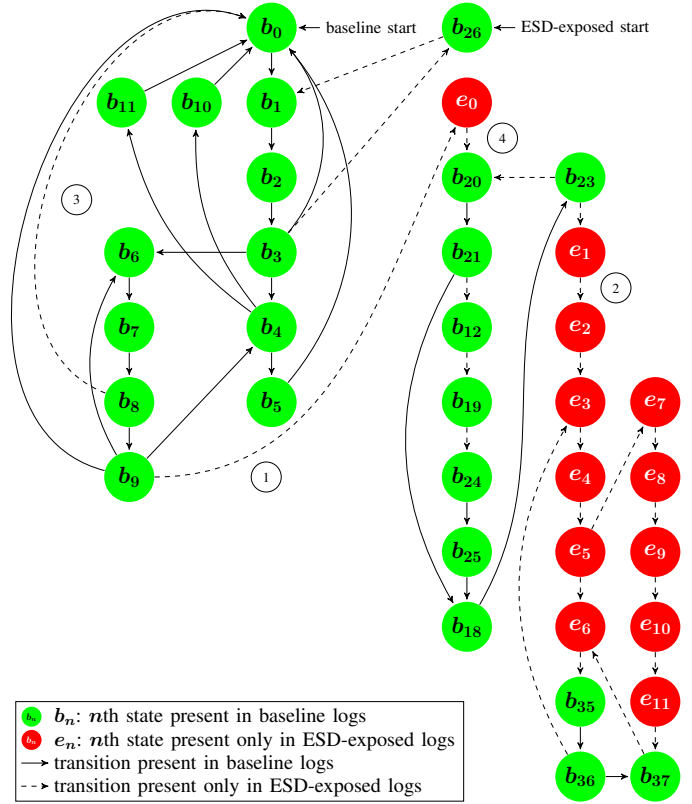
\begin{figure}
	\centering
%
%
%
%


\newenvironment{customlegend}[1][]{%
    \begingroup
    \csname pgfplots@init@cleared@structures\endcsname
    \pgfplotsset{#1}%
}{%
    \csname pgfplots@createlegend\endcsname
    \endgroup
}%

\def\addlegendimage{\csname pgfplots@addlegendimage\endcsname}%


\pgfkeys{/pgfplots/number in legend/.style={%
        /pgfplots/legend image code/.code={%
            \node at (0.125,-0.0225){#1}; 
        },%
    },
}%
\pgfplotsset{%
every legend to name picture/.style={west}%
}%

\vspace*{-4ex}%
\hspace*{-40pt}%
\scalebox{0.64} {%
\begin{tikzpicture}[->,>=stealth',shorten >=1pt,auto,node distance=0.5cm,
                    semithick]
  \tikzstyle{every state}=[fill=red,draw=none,text=black,font=\LARGE\boldmath,scale=0.8,minimum size=1.3cm]
	\tikzset{initial text={baseline start}}
  \node[initial right,state,fill=green] (b_0) {$b_{0}$};
\node[state,fill=green] (b_1) [below =  of b_0] {$b_{1}$};
\node[state,fill=green] (b_2) [below =  of b_1] {$b_{2}$};
\node[state,fill=green] (b_3) [below =  of b_2] {$b_{3}$};
\node[state,fill=green] (b_4) [below =  of b_3] {$b_{4}$};
\node[state,fill=green] (b_5) [below =  of b_4] {$b_{5}$};
\node[state,fill=green] (b_6) [left =  1.9cm of b_3] {$b_{6}$};
\node[state,fill=green] (b_7) [below =  of b_6] {$b_{7}$};
\node[state,fill=green] (b_8) [below =  of b_7] {$b_{8}$};
\node[state,fill=green] (b_9) [below =  of b_8] {$b_{9}$};
\node[state,fill=green] (b_10) [left =  of b_1] {$b_{10}$};
\node[state,fill=green] (b_11) [left =  of b_10] {$b_{11}$};

\tikzset{initial text={ESD-exposed start}}
\node[initial right,state,fill=green] (b_26) [right =  3cm of b_0] {$b_{26}$};
\node[state,fill=red,text=white] (e_0) [below =  of b_26] {$e_{0}$};
\node[state,fill=green] (b_20) [below =  of e_0] {$b_{20}$};
\node[state,fill=green] (b_21) [below =  of b_20] {$b_{21}$};
\node[state,fill=green] (b_12) [below =  of b_21] {$b_{12}$};
\node[state,fill=green] (b_19) [below =  of b_12] {$b_{19}$};
\node[state,fill=green] (b_24) [below =  of b_19] {$b_{24}$};
\node[state,fill=green] (b_25) [below =  of b_24] {$b_{25}$};
\node[state,fill=green] (b_18) [below =  of b_25] {$b_{18}$};
\node[state,fill=green] (b_23) [right = 1.3cm of b_20] {$b_{23}$};
\node[state,fill=red,text=white] (e_1) [below =  of b_23] {$e_{1}$};
\node[state,fill=red,text=white] (e_2) [below =  of e_1] {$e_{2}$};
\node[state,fill=red,text=white] (e_3) [below =  of e_2] {$e_{3}$};
\node[state,fill=red,text=white] (e_4) [below =  of e_3] {$e_{4}$};
\node[state,fill=red,text=white] (e_5) [below =  of e_4] {$e_{5}$};
\node[state,fill=red,text=white] (e_6) [below =  of e_5] {$e_{6}$};
\node[state,fill=green] (b_35) [below =  of e_6] {$b_{35}$};
\node[state,fill=green] (b_36) [below =  of b_35] {$b_{36}$};
\node[state,fill=red,text=white] (e_7) [right =  of e_3] {$e_{7}$};
\node[state,fill=red,text=white] (e_8) [below =  of e_7] {$e_{8}$};
\node[state,fill=red,text=white] (e_9) [below =  of e_8] {$e_{9}$};
\node[state,fill=red,text=white] (e_10) [below =  of e_9] {$e_{10}$};
\node[state,fill=red,text=white] (e_11) [below =  of e_10] {$e_{11}$};
\node[state,fill=green] (b_37) [below =  of e_11] {$b_{37}$};

\node[state,fill=none,draw=black,scale=0.6] (c1) [right = 2 of b_9] {1};
\node[state,fill=none,draw=black,scale=0.6] (c2) [below right = 0.2 of e_1] {2};
\node[state,fill=none,draw=black,scale=0.6] (c3) [above left = 0.7 of b_6] {3};
\node[state,fill=none,draw=black,scale=0.6] (c4) [below right = 0.2 of e_0] {4};

\path(e_3) edge [dashed] node {} (e_4)
(b_8) edge [bend left=90,looseness=1.1,dashed] node {} (b_0)
(e_5) edge [dashed] node {} (e_6)
(e_7) edge [dashed] node {} (e_8)
(b_9) edge [bend left = 93,looseness=1.1] node {} (b_0)
(b_36) edge [bend left,dashed] node {} (e_3)
(b_6) edge [] node {} (b_7)
(b_21) edge [bend right,] node {} (b_18)
(b_23) edge [dashed] node {} (b_20)
(b_3) edge [bend right=40,] node {} (b_0)
(e_0) edge [dashed] node {} (b_20)
(b_10) edge  node {} (b_0)
(b_21) edge [dashed] node {} (b_12)
(b_35) edge [] node {} (b_36)
(b_36) edge [] node {} (b_37)
(b_4) edge [bend left=20] node {} (b_10)
(b_5) edge [bend right=40,] node {} (b_0)
(b_4) edge [] node {} (b_5)
(b_19) edge [dashed] node {} (b_24)
(b_2) edge [] node {} (b_3)
(b_23) edge [dashed] node {} (e_1)
(b_3) edge [dashed] node {} (b_26)
(b_9) edge [bend left,] node {} (b_6)
(b_0) edge [] node {} (b_1)
(b_26) edge [dashed] node {} (b_1)
(e_10) edge [dashed] node {} (e_11)
(b_12) edge [dashed] node {} (b_19)
(e_9) edge [dashed] node {} (e_10)
(b_3) edge [] node {} (b_4)
(e_5) edge [dashed] node {} (e_7)
(b_8) edge [] node {} (b_9)
(b_9) edge  node {} (b_4)
(b_1) edge [] node {} (b_2)
(b_24) edge [] node {} (b_25)
(b_20) edge [] node {} (b_21)
(e_11) edge [dashed] node {} (b_37)
(b_3) edge  node {} (b_6)
(e_2) edge [dashed] node {} (e_3)
(b_18) edge [out=60,in=240] node {} (b_23)
(e_4) edge [dashed] node {} (e_5)
(e_6) edge [dashed] node {} (b_35)
(b_9) edge [out=0,in=245,dashed] node {} (e_0)
(b_4) edge [bend left=20] node {} (b_11)
(e_8) edge [dashed] node {} (e_9)
(b_11) edge  node {} (b_0)
(b_37) edge [dashed] node {} (e_6)
(e_1) edge [dashed] node {} (e_2)
(b_7) edge [] node {} (b_8)
(b_25) edge [] node {} (b_18)
;

\begin{customlegend}[legend cell align=left, 
legend entries={
$b_n$: $n$th state present in baseline logs,
$e_n$: $n$th state present only in ESD-exposed logs,
transition present in baseline logs,
transition present only in ESD-exposed logs
},
legend style={at={(4,-14)},font=\large\boldmath}] 
\addlegendimage{legend image code/.code={\node [state,fill=green,scale=0.4,font=\LARGE\boldmath] {$b_n$};}}
\addlegendimage{legend image code/.code={\node [state,fill=red,text=white,scale=0.4,font=\LARGE\boldmath] {$b_n$};}}
		\addlegendimage{}
		\addlegendimage{dashed}
\end{customlegend}
\end{tikzpicture}
}
\vspace*{-1.8ex}%
%
%
	\caption{Execution graph of one baseline trace and one ESD-exposed execution trace}
	\label{fig:state-graph}
\end{figure}

Figure~\ref{fig:state-graph} shows an execution graph of sample baseline and ESD-exposed execution traces.
The set of nodes and solid arcs on the left of the figure is the execution graph of the baseline log.
The right of the figure consists of the additional states and transitions present in the sample ESD-exposed log. 
This execution graph demonstrates several potential effects of ESD on the system state: \circled{1} transitions to non-baseline states, \circled{2} transitions between non-baseline states, \circled{3} non-baseline transitions between baseline states, and \circled{4} transitions from non-baseline to baseline states. 

\begin{figure}
  \centering
  \includegraphics[width=\columnwidth]{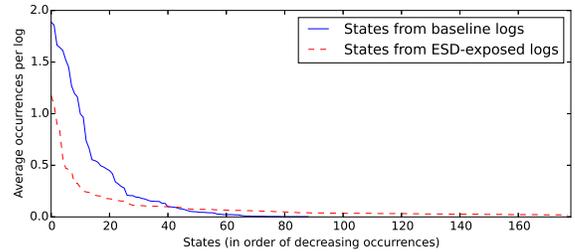}
  \caption{Average State Occurrences Per Log} 
  \label{fig:histogram}
\end{figure}

Consider how we should expect the system to behave under normal conditions and under ESD exposure.
Normally, it should have a small number of common code paths and some edge case handling.
Under ESD exposure, we should see a number of anomalous states caused by various register bits being flipped as well as control flow anomalies.
Figure~\ref{fig:histogram} shows the average number of occurrences of states per baseline and ESD-exposed traces.
The baseline traces show a few states that are very common and a small tail of less common states.
There are far more unique states in ESD-exposed logs, and they are far less likely to occur.
(We have omitted half of the ESD-exposed state tail to make the interesting portion of the graph more legible.)
This graph provides quick verification of our methodology; we can see that the data we have collected reflects expected system behavior.

\begin{figure}
\centering
\includegraphics[width=\columnwidth]{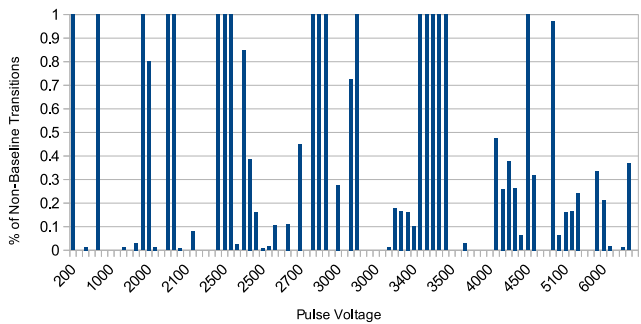}
\caption{Relationship between pulse voltage and ESD-caused transitions}
\label{fig:tranState}
\end{figure}

Figure~\ref{fig:tranState} compares the TLP pulse voltage with the percentage of transitions to or from states not in the baseline logs.
The lack of a clear relationship between observed ESD coupling and pulse voltage indicates that there are confounding factors between ESD exposure and system behavior.
These factors may include field type and orientation, injection location, pulse frequency, and the operation being performed by the host controller at the time of injection.
In addition, the ESD injection may cause the system to crash almost instantaneously, in which case the resulting state log will have relatively few states caused by ESD\@.
More work is needed to assess the effect each of these factors has on system operation. 

\section{Conclusion}
\label{sec:conclusion}

We have presented a software-based methodology for detecting ESD events on embedded system peripherals. 
This methodology monitors the state of the peripheral by reading the registers it exposes to the CPU with an instrumented kernel driver for the peripheral.
As with all software monitoring techniques, this approach is only able to monitor events that do not entirely disrupt CPU execution.
We applied this methodology to a USB host controller on an embedded system running Linux.
We demonstrated that we are able to observe states and transitions that the system experiences only when exposed to ESD\@.

Furthermore, the relationship between the recorded errors and {ESD} can be reversed.
Doing so allows us to predict, based on the errors that the software experiences, when and where the system experiences {ESD}.
We can apply this in several ways: components that have received ESD can be identified, either for replacement (if the goal of the experiment is to repair hardware that has experienced ESD) or for improvement (if the goal is to reduce the effects of {ESD} on a peripheral);
in addition, software can be written to recover from error states in a more efficient and automatic fashion. 
Software may also be able to compensate for the effects of {ESD}, allowing operation to continue in hostile environments at the cost of reduced performance and more software overhead.


A topic for future research is correlating system states with ESD injection on a specific location on the board, which could give insight into which components have experienced ESD for performing repairs or assist circuit designers in shielding the board from particular error states.
One could also study system states from a software perspective to determine how best to recover from certain ESD-induced errors. 
Finally, applying this methodology to other peripherals and embedded systems may lead to additional insights for software monitoring. 
In particular, applying this methodology in tandem with PCB schematic and chip layout analysis would provide a bridge between software-observed and hardware-observed ESD effects.


\bibliographystyle{IEEEtranN}
\bibliography{IEEE_bibstyle,biblio}

\end{document}